\documentclass[aps,prl,twocolumn,superscriptaddress,nofootinbib]{revtex4-2}
\usepackage[english]{babel}
\usepackage{bm}
\usepackage{pdfpages} 
\usepackage{pgffor} 
\usepackage{amsmath}
\usepackage{amssymb}
\usepackage{dutchcal}
\usepackage[colorlinks=true, allcolors=cyan]{hyperref}
\usepackage{diagbox}
\usepackage{multirow}
\usepackage{threeparttable}
\usepackage{xfrac}
\usepackage{xcolor}

\makeatletter
\AtBeginDocument{\let\LS@rot\@undefined}
\makeatother
\def\supplementfilename{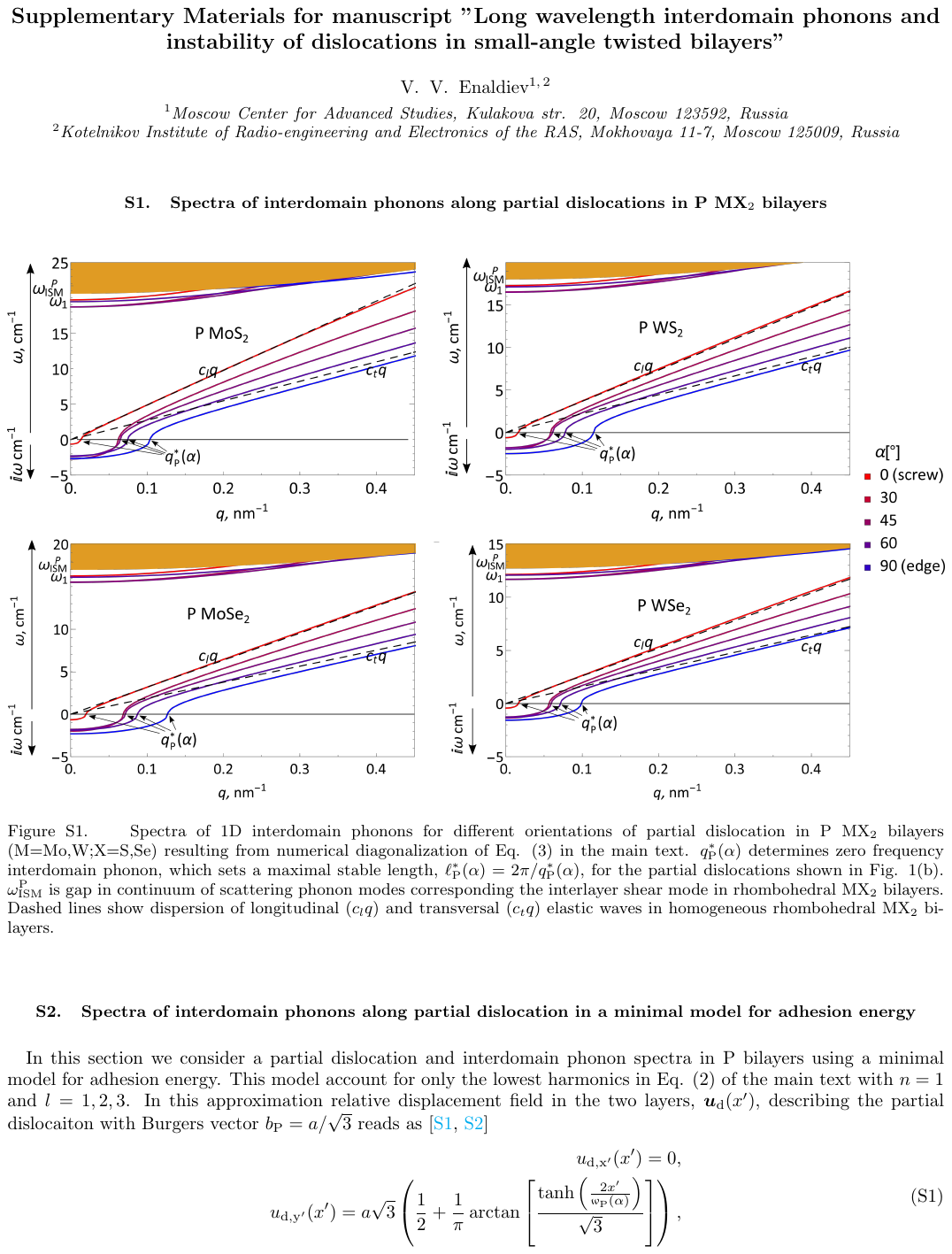}
\pdfximage{\supplementfilename}
\def\numbersupplementpages{\the\pdflastximagepages}
\newif\ifarXiv
\arXivtrue

\begin{document}
\title{Long wavelength interdomain phonons and instability of dislocations in small-angle twisted bilayers}

\author{V.~V.~Enaldiev}
\affiliation{Moscow Center for Advanced Studies, Kulakova str. 20, Moscow 123592, Russia} 
\affiliation{Kotelnikov Institute of Radio-engineering and Electronics of the RAS, Mokhovaya 11-7, Moscow 125009, Russia}
\email{vova.enaldiev@gmail.com}

\begin{abstract}
We develop a theory for long wavelength phonons originating at dislocations separating domains in small-angle twisted homobilayers of 2D materials such as graphene and MX$_2$ transition metal dichalcogenides (M=Mo,W; X=S,Se). We find that both partial and perfect dislocations, forming due to lattice relaxation in the twisted bilayers with parallel and anti-parallel alignment of unit cells of the constituent layers, respectively, support several one-dimensional subbands of the {\it interdomain} phonons. We show that spectrum of the lowest gapless subband is characterized by imaginary frequencies, for wave-numbers below a critical value, dependent on the dislocation orientation, which indicates an instability for long enough straight partial and perfect dislocations. We argue that pinning potential and/or small deformations of the dislocations could stabilize the gapless phonon spectra. The other subbands are gapped, with subband bottoms lying below the frequency of interlayer shear mode in domains, which facilitates their detection with the help of optical and magnetotransport techniques.
\end{abstract}

\maketitle

Moir\'e superlattices (mSL), emerging at the interface of vertically assembled two-dimensional materials with commensurate lattice parameters and controllable interlayer twist, have enriched physics of van der Waals (vdW) heterostructures. This includes formation of flat minibands in twisted graphene \cite{Bistritzer2011,AitorPRB2021} and transition metal dichalcogenide \cite{Zhang2020,Naik2020,Magorrian2021,Angeli2021} (TMD) bilayers, responsible for strongly correlated insulating electron phases \cite{Kennes2021,Foutty2023,Campbell2024,Li2021a} and superconductivity \cite{cao2018unconventional,Yankowitz2019}, as well as interfacial ferroelectricity recently discovered in small-angle twisted TMD \cite{Weston2022,Wang2022} and hexagonal boron nitride \cite{woods2021,yasuda2021,stern2021} bilayers. The latter shows up due to adhesion-promoted lattice relaxation of the layers, slightly misoriented from parallel (P) orientation of unit cells across interface, which transforms mSL into array of triangular rhombohedral polar domains, separated by network of partial dislocations. Such a structural transition is generic for long-period mSL, specific for small-angle twisted homobilayers, observed also in graphene \cite{yoo2019atomic,AldenPNAS} and TMD \cite{Weston2020,rosenberger2020,Halbertal2021,Liang2023} bilayers with antiparallel (AP) orinetation of unit cells. Similarly to P TMD, relaxed mSL for twisted bilayer graphene feature triangular domains with Bernal stacking adjoining each other via network of partial dislocations, whereas for AP TMD, relaxed mSL comprises hexagonal
2H-stacked domains separated by perfect dislocations. 

\begin{figure}[!t]
	\includegraphics[width=1.0\columnwidth]{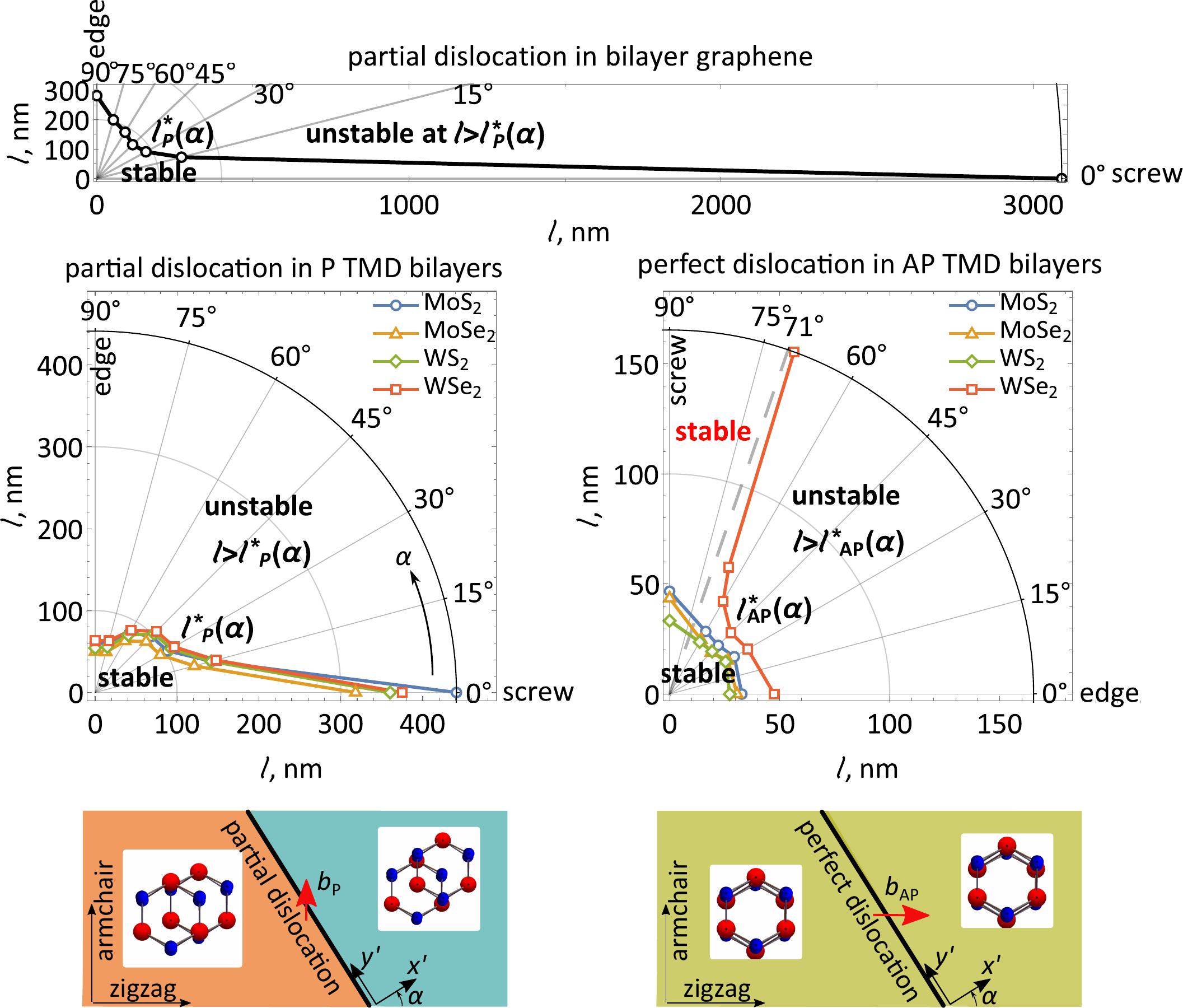}
	\caption{\label{fig:1} (a,b) Polar plots demonstrate regions of (in)stability for straight partial dislocations in graphene and P TMD bilayers in parametric space of dislocations' length, $\ell$, and orientation $\alpha$ with respect to Burgers vector $\bm{b}_{\rm P}$ ($|b_{\rm P}|=a/\sqrt{3}$). For each material, boundary between the regions is set by $\ell^*_{\rm P}(\alpha)=2\pi/q^*_{\rm P}(\alpha)$, where $q^*_{\rm P}(\alpha)$ determines zero frequency mode of the interdomain phonons, shown in Fig. \ref{fig:1DspectraP},\ref{fig:3}. (c) Same for perfect dislocations in AP TMD bilayers, with $\alpha$ counted from normal to $\bm{b}_{\rm AP}$ ($|\bm{b}_{\rm AP}|=a$). For AP WSe$_2$ bilayers (red) perfect dislocations are stable for every lengths within interval $|\alpha^{\circ}-90^{\circ}|<19^{\circ}$ around screw orientation  ($\alpha^{\circ}=90^{\circ}$), which occurs due to absence of imaginary frequencies in interdomain phonon spectra (see Fig. \ref{fig:3}). Bottom insets show orientation of dislocations with respect to crystallographic axes as well as $x'Oy'$ reference frame  used in Eq. \eqref{Eq:2} } 
\end{figure}

Effect of spatial modulation of interlayer adhesion on lattice dynamics of twisted heterostructures has recently been studied experimentally \cite{Quan2021,Jong2022} and theoretically  \cite{BalandinPRB2013,KoshinoPRB2019,PRB2019Ochoa,Maity2020,PRB2022_Samajdar,Liu2022,Jong2022,Lu2022,PRB2023Cappelluti,PRB2023Girotto} with a main focus on twist-angle dependence of phonon spectra. However, for long-period mSL, composed of domains and networks of dislocations \cite{yoo2019atomic,AldenPNAS,Weston2020,rosenberger2020,Halbertal2021}, periodicity is usually broken \cite{Butz2013,Molino2023,Liang2023,Engelke2023} by unintentional strain introduced on sample manufacturing stage. This poses a more relevant question on lattice vibrations in the bilayers with macroscopic domains adjoining each other via a single dislocation.

In this Letter we develop a theory for phonons, called below {\it interdomain}, that are formed at and freely propagating along a single dislocation (partial for P TMD and graphene, and perfect for AP TMD) separating domains in the small-angle twisted bilayers. Analyzing one-dimensional (1D) spectra of the interdomain phonons, characterized by longitudinal wave-number $q$, we find that the lowest gapless subband have imaginary frequencies at $q<q^*_{\rm P/AP}(\alpha)$, where $q^*_{\rm P/AP}(\alpha)$ is zero-frequency-mode wave-number, determined by the dislocation orientation, set by angle $\alpha$. Formation of the modes with imaginary frequencies indicates an instability of straight dislocations with the lengths, $\ell$, exceeding the critical value, $\ell^*_{\rm P/AP}(\alpha)=2\pi/q^*_{\rm P/AP}$, depicted on polar plots in Fig. \ref{fig:1}. The critical lengths are finite for partial/perfect dislocations in all studied bilayers except AP WSe$_2$, where the perfect dislocations with orientation close to screw (within $19^{\circ}$) are stable for all range of lengths due to the gapped spectra of the interdomain phonons. In addition, we find that partial/perfect dislocations support at least one gapped subband with the subband bottom under frequency of interlayer shear mode in domains, enabling their manifestation in optical and magnetotransport experiments.

The developed theory of the interdomain phonons exploits a continuum approach justified for description of long-wavelength lattice vibrations \cite{KoshinoPRB2019,PRB2019Ochoa,Enaldiev_PRL,PRB2022_Samajdar,PRB2023Girotto}. For long-wavelength phonons in P/AP MX$_2$ and graphene bilayers with a single dislocation Lagrange function reads as 
\begin{multline}\label{Eq:1}
	\mathcal{L}=
	\sum_{l=t,b}\left[\dfrac{\rho\dot{\bm{u}}^{2}_{l}}{2} - \dfrac{\lambda}{2}\left(u_{ii}^{(l)}\right)^2-\mu \left(u^{(l)}_{ij}\right)^2\right]\\
	 - W_{\rm ad}(\bm{u}_{t}-\bm{u}_{b}). 
\end{multline}  
Here, terms in square brackets describe kinetic and elastic energies of deformations, \mbox{$\bm{u}^{t/b}(\bm{r},t)=\bm{u}^{t/b}_{\rm d}(\bm{r})+\bm{u}^{t/b}_{\rm w}(\bm{r},t)$} in top (t)/bottom (b) layers, with $\bm{u}_{\rm d}$ and $\bm{u}_{\rm w}$ characterizing dislocation and phonon displacements, respectively, and \mbox{$2u^{ (t/b)}_{ij}=\partial_iu^{(t/b)}_j+\partial_ju^{(t/b)}_i,$} is in-plane strain ($ i,j=x,y$); $\rho$ is areal mass density of monolayer and $\lambda$ and $\mu$ are its Lam\'e elastic parameters. The last term in Eq. \eqref{Eq:1} stands for interlayer adhesion energy density of graphene and P/AP TMD bilayers \cite{Zhou2015,CarrPRB2018,Enaldiev_PRL,Enaldiev2024}:  
\begin{align}\label{Eq:adhesion}
	W_{\rm ad}(\bm{r}_0) =&\sum_{\substack{n,l=1,2,3}}W_{l}^{(n)}(\bm{r}_0),\\
	W_{l}^{(n)}(\bm{r}_0)=&w^{(s)}_{n}\cos\left(\bm{G}^{(n)}_l\bm{r}_0\right) + w^{(a)}_{n}\sin\left(\bm{G}^{(n)}_l\bm{r}_0\right). \nonumber
\end{align}
Here, \mbox{$\bm{r}_0$} is an in-plane lateral offset between top and bottom layers counted from AA stacking (\mbox{$\bm{r}_0^{\rm AA}=0$}) for P and AP alignments of layers, $\bm{G}^{(n)}_{l}=\sqrt{n+1-\delta_{n,1}}\hat{R}_{\frac{2\pi(l-1)}{3}+\frac{\pi}{2}\delta_{n,2}}\bm{G}$ are three triads of reciprocal lattice vectors determined by an anti-clockwise $\varphi$-rotation, $\hat{R}_{\varphi}$, around out-of-plane axis, reciprocal basis vector, $\bm{G}=\frac{4\pi}{a\sqrt{3}}(\sqrt{3}/2,1/2)$ ($a$ is monolayer lattice parameter), and Kronecker delta $\delta_{i,j}$. Values of $w^{(s,a)}_{1,2,3}$, $\lambda$ and $\mu$ for studied materials can be found in Refs. \cite{Enaldiev2024,Zhou2015,CarrPRB2018}. 
  
We note that interlayer adhesion \eqref{Eq:1} depends only on relative displacements of atoms in t/b layers, so that center of mass and relative motions of the layers' unit cells are decoupled. Below, we study only layer-antisymmetric shear phonons, $\bm{u}_{\rm w}=\bm{u}^{t}_{\rm w}-\bm{u}^{b}_{\rm w}$, which are influenced by interlayer adhesion, whereas layer-symmetric in-plane displacements does not suffer from dislocation formation. 

Since partial dislocations in small-angle twisted P TMD and graphene bilayers can be characterized by the same Burgers vector \cite{AldenPNAS,Butz2013,Annevelink2020,Enaldiev_PRL,Enaldiev2024,Engelke2023}, $b_{\rm P}=a/\sqrt{3}$, we call the structures as P bilayers. In P bilayers the partial dislocations provide a transition of interlayer stacking register between mirror-twinned domains -- rhombohedral \cite{Weston2020,rosenberger2020,Liang2023} for MX$_2$ and Bernal \cite{AldenPNAS,yoo2019atomic} for graphene. In small-angle twisted AP TMD bilayers (AP bilayers) boundaries of 2H-stacked domains are perfect dislocations \cite{Weston2020,rosenberger2020}, characterized by Burgers vector \cite{Weston2020,rosenberger2020}, $b_{\rm AP}=a$, describing  lattice translation on a full period, $b_{\rm AP}=a$.

\begin{table}
	
	\caption{Frequencies (in cm$^{-1}$) of interdomain phonon subband bottoms, $\omega_{n}$, for screw dislocations and interlayer shear mode in rhombohedral/2H and Bernal domains in P/AP TMD and graphene bilayers, respectively. \label{tab:omega1}}
	\resizebox{\columnwidth}{!}{
		\begin{threeparttable}
			\begin{tabular}{lc|c|c|c|c|c}
				\hline
				\hline
				&  & Gr  &   MoS$_2$ &  WS$_2$ &  MoSe$_2$ &  WSe$_2$  \\ 
				\hline 
				\hline
				&&\multicolumn{5}{c}{P bilayers}\\
				\hline 
				\hline
				\multirow{2}{*}{$\omega_{\rm ISM}^{\rm P}$} &  this work & 33.8 & 20.7 & 18.1 & 17.0 & 12.8  \\
				&  exp. & 32 \cite{Tan2012} & 23\cite{Quan2021} & 18\cite{Zhang2024} & 18 \cite{Brien2016} &   \\
				\hline
				$\omega_1$& & 32.4 & 19.7 & 17.3 & 16.3 & 12.2  \\ 
				\hline
				\hline
				&&&\multicolumn{4}{c}{AP bilayers}\\
				\hline
				\hline
				\multirow{2}{*}{$\omega_{\rm ISM}^{\rm AP}$} & this work & NA & 22.5 & 18.1 & 18.8 & 14.3  \\
				&  exp. & & 22 \cite{Zhao2013,Lee2017} & 17.6\cite{Boora2024}, 18\cite{Zhang2024} & 18\cite{Brien2016} & 17\cite{Zhao2013} \\
				\hline
				$\omega_0$ & & NA & 0 & 0 & 0 & 2.7  \\ 
				\hline
				$\omega_1$ & & NA & 12.3 & 9.5 & 10.9 & 11.6  \\
				\hline
				$\omega_2$ & & NA & 21.2 & 17.8 & 17.9 & -  \\
				\hline
				\hline
			\end{tabular}
		\end{threeparttable}
	}
\end{table}

To be specific, we consider P (AP) bilayers having a single partial (perfect) dislocation characterized by $\bm{b}_{\rm P}=(0,a/\sqrt{3})$ ($\bm{b}_{\rm AP}=(a,0)$) and orientation set by a normal $\bm{n}(\alpha)=(\cos(\alpha),\sin(\alpha))$, in a fixed Cartesian coordinates with $x$-axis ($y$-axis) along the zigzag (armchair) crystallographic direction. In this case $\alpha=0$ corresponds to screw (edge) dislocation orientation for P (AP) bilayers. In absence of phonon displacements (i.e. $\bm{u}_{\rm w}=0$), minimization of Lagrangian \eqref{Eq:1} gives rise to dislocation profile $\bm{u}_{\rm d}$ that was previously determined in Refs. \cite{Enaldiev_PRL,Enaldiev2024}. Hereafter we consider $\bm{u}_{\rm d}$ as a known function of coordinates across dislocation line. Taking advantage of translational invariance of straight dislocations along its length, we use a rotated Cartesian coordinates with $x'$-axis ($y'$-axis) along the unit vector $\bm{n}(\alpha)$ in P (AP) bilayers. Then, taking into account phonons in Lagrangian \eqref{Eq:1} and minimizing it on $\bm{u}_{\rm w}$, we find linearized Lagrange-Euler equations as follows:  
\begin{multline}\label{Eq:2}
	\begin{pmatrix}
		-c_l^2\partial_{x'}^2-c_t^2\partial^2_{y'}-\omega^2 & -(c_t^2-c^2_l)\partial_{x'}\partial_{y'} \\
		-(c_t^2-c^2_l)\partial_{x'}\partial_{y'} & -c_t^2\partial_{x'}^2-c_l^2\partial^2_{y'}-\omega^2
	\end{pmatrix}
	\begin{pmatrix}
	u_{{\rm w},x'} \\
		u_{{\rm w},y'}
	\end{pmatrix}
	= \\
	\tfrac{16\pi^2}{3a^2\rho}\sum_{n,l=1,2,3}(n+1-\delta_{n,1})W^{(n)}_{l}(\bm{u}_{\rm d}(\bm{r}'))\times \\
	\begin{pmatrix}
		2\cos^2\left(\alpha_{l,n}\right) & -\sin\left(2\alpha_{l,n}\right) \\
		-\sin\left(2\alpha_{l,n}\right) & 2\sin^2\left(\alpha_{l,n}\right)
	\end{pmatrix}
	\begin{pmatrix}
		u_{{\rm w},x'} \\
		u_{{\rm w},y'}
	\end{pmatrix}.
\end{multline}
Here, $c_{t}=\sqrt{\mu/\rho}$ and $c_{l}=\sqrt{(\lambda+2\mu)/\rho}$ are speed of transverse and longitudinal elastic waves of isolated layers, $\partial_{x'}$, $\partial_{y'}$ are derivatives on the rotated coordinates $x',y'$ and $\alpha_{l,n}=\alpha+\pi\left(\tfrac{l}{3}+\tfrac{n}{2}\right)$. Below, we analyze spectra of 1D interdomains phonons for a single partial/perfect dislocation in P/AP bilayers, given by  Eq. \eqref{Eq:2}. 

First, we consider P bilayers substituting in Eq. \eqref{Eq:2} displacement field of a partial dislocation $u_{\rm d}(x')$. Far from the dislocation core $x'=0$ (i.e. at $\bm{u}_{\rm d}(\pm\infty)=(3\pm1)a/2\sqrt{3}$) Eq. \eqref{Eq:2} describes continuum of scattering states characterized by a spectral gap, $\omega_{\rm ISM}^{\rm P}=\sqrt{\tfrac{\pi^2}{3\rho a^2}\left(24w_1^{(s)}-144w_2^{(s)}+128w_3^{(s)}\right)}$, corresponding to frequency of interlayer shear mode (ISM) of Bernal/rhombohedral-stacked domain in graphene/TMD bilayer. In Table \ref{tab:omega1} we list values of the gaps for studied materials and compare with those extracted from Raman scattering experiments.  

Using homogeneity of Eq. \eqref{Eq:2} along the partial dislocation we characterize interdomain phonons by longitudinal wave-number, $q$: $\bm{u}_{\rm w}=(u^{(q)}_{x'},u^{(q)}_{y'})e^{i\left(qy'-\omega t\right)}$. To determine amplitudes and dispersion of the interdomain phonons we use method of finite differences. In particular, we introduce a grid $x_j=jh-L/2$ set in the interval $-L/2\leq x_j\leq L/2$ with a step \mbox{$h\ll \mathcal{w}$} and $L\gg\mathcal{w}$ ($\mathcal{w}$ is the partial dislocation width), making the following substitutions in Eq. \eqref{Eq:2}: $\partial^2_x \bm{u}^{(q)}\to (\bm{u}^{(q)}_{j+1}-2\bm{u}^{(q)}_{j}+\bm{u}^{(q)}_{j-1})/h^2$,  $\partial_x \bm{u}^{(q)}\to (\bm{u}^{(q)}_{j+1}-\bm{u}^{(q)}_{j-1})/2h$, and $\partial_y \bm{u}^{(q)}\to iq\bm{u}^{(q)}_j$,  where \mbox{$\bm{u}^{(q)}_{j}\equiv\bm{u}^{(q)}(x_{j})$}. Afterwards, we numerically diagonalize corresponding matrix for amplitudes $\bm{u}^{(q)}_j$ at every $q$.

\begin{figure}[!h]
	\includegraphics[width=1.0\columnwidth]{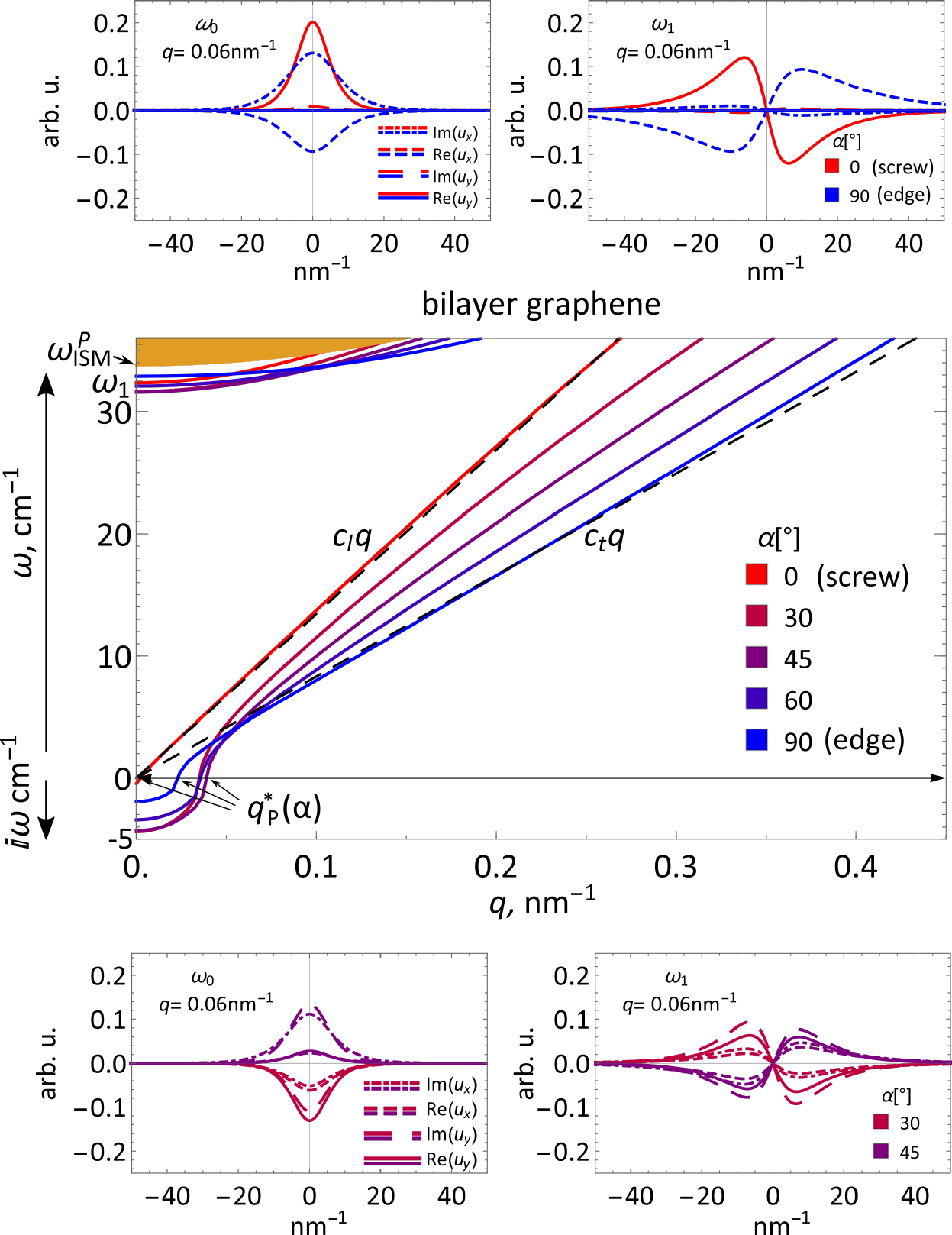}
	\caption{(a) Spectra of 1D interdomain phonons for different orientations of partial dislocation in bilayer graphene. For every $\alpha$, $q_{\rm P}^{*}(\alpha)$ determines boundary of instability regions for partial dislocations, shown in Fig. \ref{fig:1}(a,b). $\omega_{\rm ISM}^{\rm P}\approx34\,{\rm cm}^{-1}$ is gap in scattering continuum corresponding to interlayer shear phonon mode in Bernal bilayer graphene. Dashed lines show dispersion of longitudinal ($c_lq$) and transversal ($c_tq$) sound waves in Bernal bilayer graphene. Top and bottom insets show distributions of displacements of interdomain phonons belonging to gapless ($\omega_0$) and gapped ($\omega_1$) subbands. \label{fig:1DspectraP}}
\end{figure}

In Fig. \ref{fig:1DspectraP} we demonstrate resulting spectra for a set of partial dislocation orientations in bilayer graphene. In general, two (gapless and gapped) 1D subbands of interdomain phonons are formed under scattering state continuum for every orientation of partial dislocation. This happens for other materials as well (see SI). 

Unexpectedly, for every orientation ($\alpha$) of a partial dislocation the gapless subband possesses purely imaginary frequencies in the long-wavelength limit at $|q|<q_{\rm P}^{*}(\alpha)$. This means that a straight partial dislocation becomes unstable when its lengths, $\ell$, overcomes a maximal value $\ell^*(\alpha)=2\pi/q_{\rm P}^{*}(\alpha)$. In $(\ell,\alpha)$-space  $\ell^*(\alpha)$-dependence sets a boundary of (in)stability region shown as a polar plot  for P bilayers in Fig. \ref{fig:1} (a,b). The smallest imaginary frequency for zero wave-number (see Fig. \ref{fig:1DspectraP}) makes screw orientation, $\alpha=0$, to be the most stable for all P bilayers, whereas deviation from the orientation strongly suppresses value of $\ell^*_{\rm P}(\alpha)$, see Fig. \ref{fig:1}(a,b). Note that it is screw partial dislocations which are formed upon lattice relaxation in ideal ($C_3$-symmetric) small-angle twisted P bilayers \cite{yoo2019atomic,Weston2020}. At $q>q^*_{\rm P}(\alpha)$ the gapless subband modes have real frequencies with group velocity following $\approx\sqrt{c_l^2\cos^2\alpha+c_t^2\sin^2\alpha}$, see Fig. \ref{fig:1DspectraP}. This agrees with polarization dependence of the modes on orientation, shown on the insets of Fig. \ref{fig:1DspectraP}. Namely, for screw/edge partial dislocations the interdomain phonons possess longitudinal/transversal polarisation of displacements, whereas mixed-type partial dislocations the polarization is elliptic.  

We note that in a minimal model for partial dislocations \cite{Lebedeva2019,Enaldiev2024} that neglects higher harmonics in adhesion energy \eqref{Eq:adhesion} ($w^{(s,a)}_{2,3}=0$), spectra of the interdomain phonons for screw and edge orientations are gapless without imaginary frequencies at any $q$, whereas for mixed-type partial dislocations the interdomain phonons still have imaginary frequencies in the long wavelength limit (see details in SI).  

The gapped 1D subband is formed just under the interlayer shear mode $\omega^{\rm P}_{\rm ISM}$ with subband bottom $\omega_{1}$ slightly dependent on $\alpha$ (see Fig. \ref{fig:1DspectraP}). In Table \ref{tab:omega1} we list values of $\omega_1$ for screw partial dislocations ($\alpha=0$) in studied bilayers, whereas dispersion of the subband, \mbox{$\omega_{1}(q)\approx \sqrt{\omega_{1}^2+c_l^2q^2\cos^2(\alpha)+c_t^2q^2\sin^2(\alpha)}$}, corresponds to polarization of displacements that varies from longitudinal to transversal for $\alpha=0$ and $\alpha=\pi/2$, respectively (see insets in Fig. \ref{fig:1DspectraP}).

Next, we consider interdomain phonons in AP TMD bilayers and substitute in Eq. \eqref{Eq:2} displacement fields $\bm{u}_{\rm d}(y')$ characterizing the perfect dislocation \cite{Enaldiev_PRL,Enaldiev2024}. As in case of partial dislocations, continuum of layer-antisymmetric scattering phonon modes possesses a spectral gap, $\omega^{\rm AP}_{\rm ISM}=\sqrt{\tfrac{8\pi^2}{\rho a^2}\left(w^{(s)}_{1}+\sqrt{3}w^{(a)}_{2}-6w_{2}^{(s)}+4w^{(s)}_{3}-4\sqrt{3}w^{(a)}_{3}\right)}$, corresponding to frequency of ISM in 2H TMD  bilayers. In Table \ref{tab:omega1} we list $\omega^{\rm AP}_{\rm ISM}$ in comparison with experimentally measured values.

\begin{figure}
	\includegraphics[width=1.0\columnwidth]{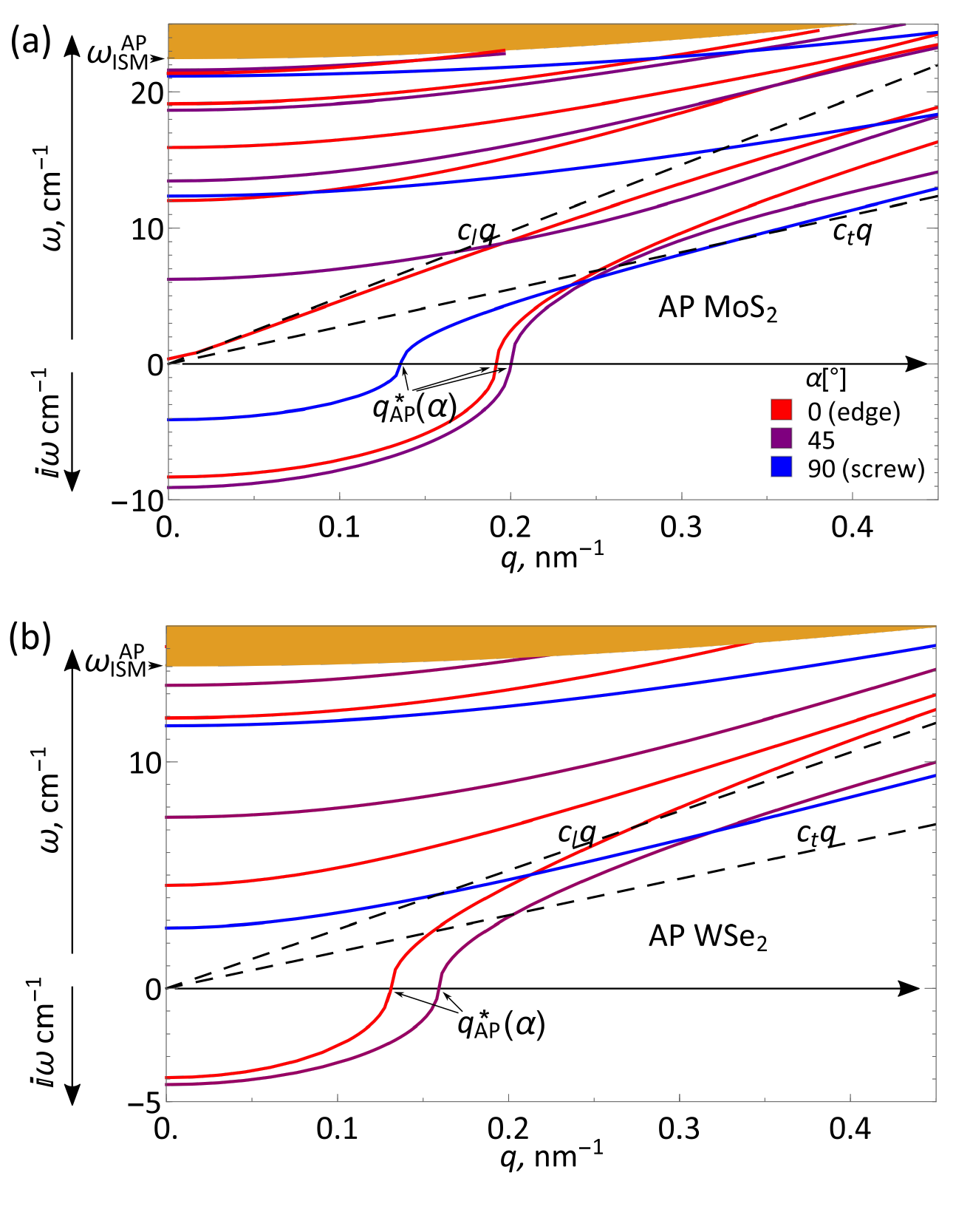}
	\caption{Panels (a) and (b) show spectra of 1D interdomain phonons for three orientations of perfect dislocation in AP MoS$_2$ and WSe$_2$ bilayers, respectively. For MoS$_2$ (MoSe$_2$ and WS$_2$, see SI) the lowest subbands possess imaginary frequencies at every $\alpha$ for $|q|<q^*_{\rm AP}(\alpha)$, indicating instability of straight perfect dislocations with length exceeding $2\pi/q^*_{\rm AP}(\alpha)$ in the bilayers. For WSe$_2$ perfect dislocations with orientation close to screw ($|\alpha^{\circ}-\leq 90^{\circ}|<19^{\circ}$) are stable as in this case subband spectra does not have imaginary frequencies at any wave numbers. Orange region in (a) and (b) shows continuum of scattering phonons. \label{fig:3}}
\end{figure}

To calculate spectra of 1D interdomain phonons in AP bilayers we use the same technique as described above for partial dislocations. In Fig. \ref{fig:3} we demonstrate spectra of interdomain phonons for AP MoS$_2$ and AP WSe$_2$ bilayers with three orientations of the perfect dislocation. Unlike the partial dislocations, two or more subbands, below scattering phonon continuum, emerge in the spectrum, which depends on dislocation orientation -- for edge dislocation the number of subbands is larger than that for screw. Moreover, the lighter material is the more subbands appear under the larger spectral gap in scattering phonon continuum (compare spectra for MoS$_2$ and WSe$_2$ in Fig. \ref{fig:3}). 

However, for AP MoS$_2$ (so as for MoSe$_2$, WS$_2$, see SI) the lowest gapless subband possesses imaginary frequencies at $|q|<q_{\rm AP}^*(\alpha)$, for every $\alpha$, whereas for AP WSe$_2$ only gapped subband are formed for screw dislocations $\alpha^\circ=90^{\circ}$, see Fig. \ref{fig:3}(b). We checked that for AP WSe$_2$ varying dislocation orientation from screw to edge leads to a gradual decrease of the lowest subband frequencies, with gapped-to-gapless transition occurring at $\alpha^{\circ}\approx71^{\circ}$ followed by formation of imaginary frequency modes at $q<q_{\rm AP}^*(\alpha)$ for smaller $\alpha$. This signifies that for AP WSe$_2$ perfect dislocations close to the screw orientation ($|\alpha^{\circ} - 90^{\circ}|<19^{\circ}$) are stable, whereas for AP MoS$_2$, AP MoSe$_2$ and AP WS$_2$ their maximal stable lengths are always finite, with the largest $\ell^*$ for screw orientation, as shown in Fig. \ref{fig:1}(c).

Spectra of the gapped subband of interdomain phonons reads as $\sqrt{\omega_n(\alpha)^2+c_l^2\cos^2\alpha+c_t^2\sin^2\alpha}$, where $\omega_{n}(\alpha)$ is the value of subband bottom, characterized by transversal quantization number $n=0,1,2,\dots$ In Table \ref{tab:omega1} we gathered values $\omega_n(\pi/2)$ for screw perfect dislocations in all studied AP bilayers. 

Formation of several subbands of interdomain phonons at partial and perfect dislocations in 2D material bilayers distinguishes them from localized phonon modes at atomically-sharp perfect dislocations in conventional 3D crystals, where it was shown \mbox{
	\cite{Ninomiya1968,Kosevich2005} }\hspace{0pt}
that the only subband of phonons localized at perfect dislocation is splitted from the continuum of scattering modes. The difference is likely results from distinction in nature of dislocations in layered van der Waals and conventional crystals. For the former they represent continuous smooth on atomic scale variation of interlayer stacking arrangement between neighboring domains, whereas, for the latter, dislocations are usually considered as extended defects with width of the order of lattice constant \cite{Litzman1965,Kosevich2005}, treated as zero width line in the elasticity theory \cite{Ninomiya1968}.

Next, we discuss possible remedies for instabilities emerging for the gapless subband. In case of partial screw dislocation, the subband features longitudinal polarization of displacements, with the highest $|\omega''|$ for $q=0$ involving uniform stretch of the dislocation. The induced strain raises energy of the gapless subband by $\mathcal{E}_{\rm P}\Delta L$, where $\mathcal{E}_{\rm P}$ is energy per unit length of partial screw dislocation and $\Delta L$ is augmentation of its length. Therefore, stability of the gapless subband is achieved when $\Delta L=\hbar\omega''/\mathcal{E}_{\rm P}$. Due to significant difference in energy scale $\hbar \omega''\sim 1 {\rm meV}$ and $\mathcal{E}_{\rm P}\sim {\rm 1 eV/nm}$ a tiny stretching is enough to ensure positive frequencies for all the modes of the gapless subband. For example, for graphene $\Delta L\approx 10^{-4}$\,nm, so that the instability is weak.

For partial edge dislocation the gapless subband is characterized by transversal polarization of interdomain phonons, with the strongest instability for the mode $q=0$ describing in-plane transverse gliding of the dislocation. In this case, stability of the dislocation could be reached by pinning potential impeding its sliding. Recently it has been observed \cite{Ko2023} that in P TMD bilayers edge disorder efficiently pins partial dislocations passing through the whole sample, whereas in relaxed moir\'e superlattices nodes of dislocation network may also inhibit the gliding instability. To add, small dislocation bending with the small curvature $\sim 2\hbar\omega''/w_{\rm P}L\mathcal{E}_{\rm P}\ll 1/L$, could help to stabilize gapless phonon spectrum at $|q|<q^*_{P}(\alpha)$, where $w_{\rm P}$ and $L$ are width and length of the partial dislocation, respectively.

For partial and perfect dislocations with both longitudinal and transverse polarization of the interdomain phonons, the largest $|\omega''|$ is also for mode $q=0$ of the gapless subband. Based on the above discussion, stability the gapless subband of interdomain phonons could be achieved by stretching of the dislocations pinned in some disorder potential.

In conclusion, one of the main results of the study is that straight partial/perfect dislocation are in general unstable in P/AP bilayers, which is due to formation of gapless subband characterized by imaginary frequencies in long wavelength part of 1D interdomain phonon spectra. It has been suggested that pinning potential and/or small deformation of a dislocation could stabilize the spectra. This finding agrees with effect of dislocation network twirling that has been recently discovered in small-angle twisted P and AP bilayers \cite{Kaliteevski2023,Jong2023,Mesple2023}. 

Furthermore, we demonstrate that dislocations in twisted bilayers also support gapped 1D subbands of interdomain phonons, lying below the frequency characterising interlayer shear mode in domains.  This could facilitate their detection with the help of optical techniques, such as Raman spectroscopy. However, formation of the interdomain phonons in sub-terahertz spectral range inflicts substantial demands on the high resolution of the Raman scattering spectra -- for AP bilayers intersubbands transition frequencies are $\approx 3-10$\,cm$^{-1}$ depending on the perfect dislocation orientation, whereas for P bilayers offset of the gapped subband bottom from the interlayer shear mode in domains is around 1\,cm$^{-1}$ (see Table \ref{tab:omega1}). Alternatively, one can take advantage of square root divergence of density of states for 1D gapped subbands, which may lead to magnetophonon resonances \cite{GurevichFirsovJETP,Parfenev1974} in 2D conductivity of small-angle twisted bilayers under quantized magnetic field. Such resonances are $1/B$-periodic, with the period equal to multiple of the gapped subband bottom frequency, which distinguishes them from Shubnikov-de Haas magnetooscillations by temperature and electron filling dependences. 

Moreover, for small-angle twisted bilayer graphene the interdomain phonons should coexist with valley-chiral \cite{Zhao2013,Ju2015,Yin2016,SanJose2014,Huang2018,Xu2019} and non-chiral \cite{barrier2024,moulsdale2024non} 1D electronic states propagating along the partial dislocations. Tuning the 1D electronic spectra by out-of-plane electric field one can promote formation of long-range charge-density-like order resulting from interaction of the interdomain phonons with the non-chiral 1D electron subbands in each valley.

{\it Acknowledgements.} The work was supported by the Russian Science Foundation (project No. 24-72-10015). Derivation of the analytical model in Supplementary Materials was supported by Ministry of Science and Higher Education of the Russian Federation (Goszadaniye). 

\bibliography{refer}
\ifarXiv


\section*{Supplementary material}

# Supplementary Materials for manuscript "Long wavelength interdomain phonons and instability of dislocations in marginally twisted bilayers"
V. V. Enaldiev[1, 2]

[1]*Moscow Center for Advanced Studies, Kulakova str. 20, Moscow 123592, Russia*
[2]*Kotelnikov Institute of Radio-engineering and Electronics of the RAS, Mokhovaya 11-7, Moscow 125009, Russia*

## S1. Spectra of interdomain phonons along partial dislocations in P MX$_2$ bilayers

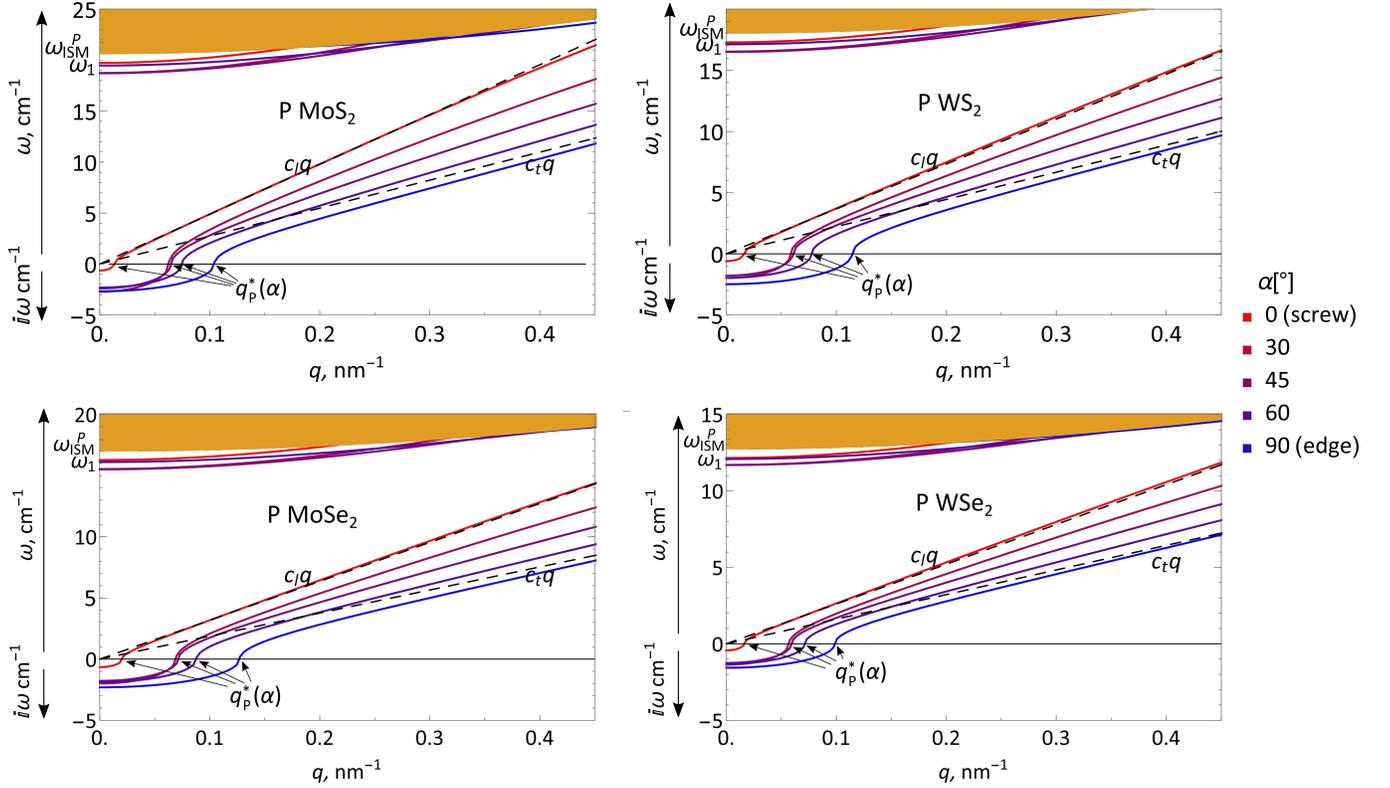

Figure S1. Spectra of 1D interdomain phonons for different orientations of partial dislocation in P MX$_2$ bilayers (M=Mo,W;X=S,Se) resulting from numerical diagonalization of Eq. (3) in the main text. $q_{\rm P}^*(\alpha)$ determines zero frequency interdomain phonon, which sets a maximal stable length, $\ell_{\rm P}^*(\alpha) = 2\pi/q_{\rm P}^*(\alpha)$, for the partial dislocations shown in Fig. 1(b). $\omega_{\rm ISM}^{\rm P}$ is gap in continuum of scattering phonon modes corresponding the interlayer shear mode in rhombohedral MX$_2$ bilayers. Dashed lines show dispersion of longitudinal ($c_l q$) and transversal ($c_t q$) elastic waves in homogeneous rhombohedral MX$_2$ bilayers.

## S2. Spectra of interdomain phonons along partial dislocation in a minimal model for adhesion energy

In this section we consider a partial dislocation and interdomain phonon spectra in P bilayers using a minimal model for adhesion energy. This model account for only the lowest harmonics in Eq. (2) of the main text with $n=1$ and $l=1,2,3$. In this approximation relative displacement field in the two layers, $\boldsymbol{u}_{\rm d}(x')$, describing the partial dislocaiton with Burgers vector $b_{\rm P}=a/\sqrt{3}$ reads as [S1, S2]

$$u_{\rm d,x'}(x') = 0,$$
$$u_{\rm d,y'}(x') = a\sqrt{3}\left(\frac{1}{2} + \frac{1}{\pi}\arctan\left[\frac{\tanh\left(\frac{2x'}{w_{\rm P}(\alpha)}\right)}{\sqrt{3}}\right]\right), \tag{S1}$$

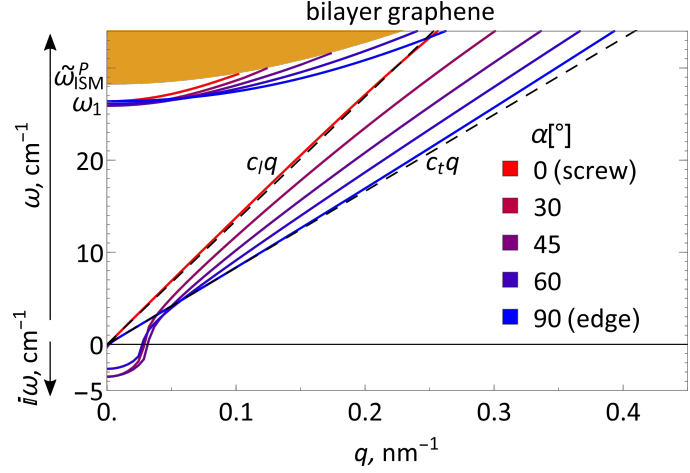

Figure S2. Spectra of 1D interdomain phonons in a minimal model for adhesion energy. $\widetilde{\omega}_{\mathrm{ISM}}^{\mathrm{P}}$ gap in scattering phonon continuum. Dashed lines show dispersion of longitudinal ($c_l q$) and transversal ($c_t q$) elastic waves in Bernal bilayer graphene.

where $w_{\mathrm{P}}(\alpha) = (2a/\pi)\sqrt{\left[\mu + (\lambda+\mu)\sin^2\alpha\right]/2w_1^{(s)}}$ is the partial dislocation width. Eq. (S1) is written in the rotated Cartesian reference frame used in the main text, with $x'$-axis perpendicular to dislocation line. Substituting Eq. (S1) and $\boldsymbol{u}_{\mathrm{w}} = (u_{x'}^{(q)}, u_{y'}^{(q)})e^{i(qy'-\omega t)}$ into Eq. (3) of the main text we obtain the following system:

$$\begin{pmatrix} c_t^2 q^2 - c_l^2 \partial_{x'}^2 - \omega^2 & (c_t^2 - c_l^2) i q \partial_{x'} \\ (c_t^2 - c_l^2) i q \partial_{x'} & c_l^2 q^2 - c_t^2 \partial_{x'}^2 - \omega^2 \end{pmatrix} \begin{pmatrix} u_{x'}^{(q)} \\ u_{y'}^{(q)} \end{pmatrix} =$$
$$\frac{16\pi^2 w_1^{(s)}}{3a^2 \rho}\left[\cos\left(\frac{2\pi u_{\mathrm{d},y'}(x')}{a\sqrt{3}}\right)\begin{pmatrix} 2+\cos(2\alpha) & -\sin(2\alpha) \\ -\sin(2\alpha) & 2-\cos(2\alpha)\end{pmatrix} + \cos\left(\frac{4\pi u_{\mathrm{d},y'}(x')}{a\sqrt{3}}\right)\begin{pmatrix} 1-\cos(2\alpha) & \sin(2\alpha) \\ \sin(2\alpha) & 1+\cos(2\alpha)\end{pmatrix}\right]\begin{pmatrix} u_{x'}^{(q)} \\ u_{y'}^{(q)} \end{pmatrix}, \quad (S2)$$

For $\omega \geq \widetilde{\omega}_{\mathrm{ISM}}^{\mathrm{P}} \equiv \sqrt{8\pi^2 w_1/\rho a^2}$ system (S2) describes continuum of scattering states, shown by orange in Fig. S2 for parameters corresponding to bilayer graphene. To find spectra of interdomain phonons in the minimal model, we first consider system (S2) for $q=0$ and screw orientation ($\alpha = 0$), for which its lines become decoupled:

$$\left[-c_l^2 \partial_{x'}^2 + V_1(x', 0)\right] u_{x'}^{(0)} = \omega^2 u_{x'}^{(0)}, \quad (S3)$$
$$\left[-c_t^2 \partial_{x'}^2 + V_2(x', 0)\right] u_{y'}^{(0)} = \omega^2 u_{y'}^{(0)}, \quad (S4)$$

where

$$V_1(x', \alpha) = \frac{16\pi^2 w_1^{(s)}}{3a^2 \rho}\frac{\left[2+\cosh\left(\frac{4x'}{w_{\mathrm{P}}(\alpha)}\right)\right]}{\left[1+2\cosh\left(\frac{4x'}{w_{\mathrm{P}}(\alpha)}\right)\right]}, \quad (S5)$$

$$V_2(x', \alpha) = \frac{16\pi^2 w_1^{(s)}}{3a^2 \rho}\left[\frac{33}{16} - \left(\frac{4+2\cosh\left(\frac{4x'}{w_{\mathrm{P}}(\alpha)}\right)}{1+2\cosh\left(\frac{4x'}{w_{\mathrm{P}}(\alpha)}\right)} - \frac{1}{4}\right)^2\right]. \quad (S6)$$

are effective potential energies. For transversal component, $u_{x'}^{(0)}$ screw dislocation represents a potential barrier, whereas for the longitudinal, $u_y^{(0)}$ – a well, for which we found an exact analytical groud state corresponding to $\omega = 0$ as

$$u_{y'}^{(0)}(x') = \frac{1}{1+2\cosh\left(\frac{4x'}{w_{\mathrm{P}}(0)}\right)}. \quad (S7)$$

Thus, for $q=0$ there is a ground mode of interdomain phonons $\boldsymbol{u}_{\mathrm{w}} = (0, u_{y'}^{(0)}(x'))$. For $q \neq 0$ the ground state form a subband with linear dispersion, $\approx c_l q$ (see Fig. S2), which were obtained by numerical solution of system (S2) using finite difference scheme, described in the main text.

Next, we consider interdomain phonons for edge partial dislocations ($\alpha = \pi/2$). As it follows from Eq. (S2), for $q = 0$, $u_{x'}^{(0)}$ and $u_{y'}^{(0)}$ are again determined by Eqs. (S3) and (S4), respectively, in which, however, one should make an exchange $V_1(x', 0) \leftrightarrow V_2(x', \pi/2)$ and $V_2(x', 0) \leftrightarrow V_1(x', \pi/2)$. As in case of the screw dislocation, we find an exact analytical solution for ground interdomain phonon mode, characterized by $\omega = 0$ but now transversal displacements $\boldsymbol{u}_{\mathrm{w}}^{(0)} = (u_{x'}^{(0)}, 0)$, with

$$u_{x'}^{(0)}(x') = \frac{1}{1 + 2\cosh\left(\frac{4x'}{w_{\mathrm{P}}(\pi/2)}\right)}.$$

For non-zero $q$ it produces a subband with group velocity of transversal sound waves $c_t$.

Thus, we have shown that in minimal model for the adhesion energy of P bilayers, the lowest subband of interdomain phonons does not manifest imaginary frequencies in spectra for screw and edge orientation of partial dislocations. However, apart from the two orientations, i.e. $\alpha \neq 0, \pi/2$ interdomain modes with imaginary frequencies still emerge in the interval $|q| < q_{\mathrm{P}}^*(\alpha)$ (see Fig. S2).

To add, the gapped subband is also formed in the minimal model with frequencies just under scattering state continuum (see Fig. S2).

## S3. SPECTRA OF INTERDOMAIN PHONONS ALONG PERFECT DISLOCATIONS IN AP MX$_2$ BILAYERS

---

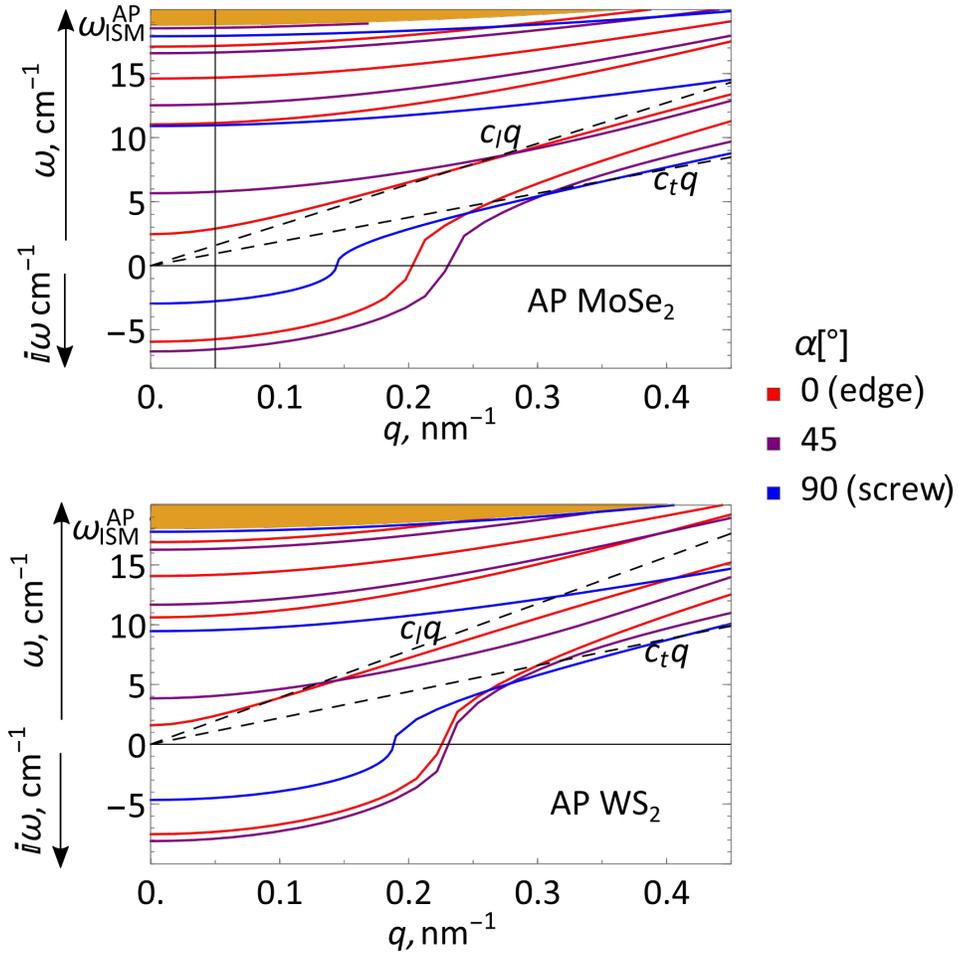

Figure S3. Spectra of 1D interdomain phonons for different orientations of perfect dislocation in AP MoSe$_2$ and AP WS$_2$ bilayers resulting from numerical diagonalization of Eq. (3) in the main text. Wave-numbers $q_P^*(\alpha)$ determines boundary of stable length for the partial dislocations, that are shown in Fig. 1(c). $\omega_{ISM}^P$ is gap in continuum of scattering phonon modes which corresponds to frequency of 2H-stacked bilayers. Dashed lines show dispersion of longitudinal ($c_l q$) and transversal ($c_t q$) elastic waves in 2H-stacked homogenuous bilayers.


\fi

\end{document}